# Single multimode fiber for *in vivo* light-field encoded nano-imaging

Zhong Wen[1,2], Zhenyu Dong[1,2], Chenlei Pang[2], Clemens F. Kaminski[3], Qilin Deng[1,2], Jinggang Xu[1,2], Liqiang Wang[1,2], Songguo Liu[2], Jianbin Tang[4], Wei Chen[5], Xu Liu[1,2,6]*, Qing Yang[2,1,6]*

[1] State Key Laboratory of Modern Optical Instrumentation, College of Optical Science and Engineering; International Research Center for Advanced Photonics, Zhejiang University, Hangzhou, 310027, China.

[2] Research Center for Humanoid Sensing, Zhejiang Lab, Hangzhou 311100, China.

[3] Department of Chemical Engineering and Biotechnology, University of Cambridge, Cambridge, CB3 0AS, United Kingdom.

[4] Center for Bionanoengineering and Key Laboratory of Biomass Chemical Engineering of Ministry of Education, College of Chemical and Biological Engineering, Zhejiang University, Hangzhou, 311100, China.

[5] Key Laboratory for Biomedical Engineering of the Ministry of Education, College of Biomedical Engineering and Instrument Science, Zhejiang University, Hangzhou, 310027, China

[6] Collaborative Innovation Center of Extreme Optics, Shanxi University, Taiyuan, 030006, China.

* Corresponding authors. Email: liuxu@zju.edu.cn (Xu Liu); qingyang@zju.edu.cn (Qing Yang)

**Super-resolution microscopy normally requiring complex and cumbersome optics is not applicable for in situ imaging through a narrow channel. Here, we demonstrate single hair-thin multimode fiber (MMF) endoscope (less than 250 μm) for *in vivo* light-field nano-imaging, which is called spatial-frequency tracking adaptive beacon light-field encoded nano-endoscopy (STABLE nano-endoscopy) that enables three-dimensional (3D) subcellular-scale imaging. Spatial-frequency tracking provides up to $10^3$ Hz disorder tracking that ensures stable imaging in long-haul MMFs (up to 200 m) under various conditions. Full-vector modulation and fluorescence emission difference are combined to enhance the imaging signal-to-noise ratio two times and to improve the resolution to sub-diffraction-limited 250 nm ($\lambda$/3NA). STABLE nano-endoscopy and white-light endoscopy (WLE) are integrated to achieve cross-scale *in vivo* imaging inside the lumen. This high-resolution and robust observation in a minimally invasive manner paves the way to gain a deeper understanding of the disease mechanisms and to bridge clinical and biological sciences.**

The real-time high-resolution endoscopy that combine imaging and therapies within the *in vivo* organ establishes an immediate endoscopic diagnosis that is virtually consistent with the histologic diagnosis, which has been the ultimate objective of endoscopists[1-2]. Recently, significant progress has been made in endoscopy. Endocytoscopy, confocal laser endomicroscopy, and other techniques have been developed to enable *in vivo* cellular imaging[3-8]. Meanwhile, super-resolution microscopy has an improved spatial resolution that is significantly smaller than the subcellular level. It leads to breakthroughs in the fields of biology and life sciences[9-13]. Despite remarkable developments in endoscopy and super-resolution microscopy, robust *in vivo* nano-endoscopy is still not satisfactorily established and rarely realized.

One promising strategy is to employ hair-thin multimode fibers (MMFs) as minimally invasive probes using wavefront shaping. Its mode densities are 2–3 orders of magnitude higher than those of the traditional endoscopes of the same diameter[14-30]. However, this technology has several intrinsic limitations. The most critical two limitations among them are the operational inflexibility due to the transmission strong dependence on its configuration[20-22] and diffraction-limited resolution with limited numerical aperture (NA)[15]. Herein, we propose an MMF based spatial-frequency tracking adaptive beacon light-field encoded (STABLE) nano-endoscope (Fig. S1). It can be used in ultra-thin (less than 250 μm) and ultra-long channels (up to 200 m). It improves the resolution to sub-diffraction-limited 250 nm (λ/3NA) with high stability and robustness against probe movement. STABLE nano-endoscopy and self-designed white-light endoscopy (WLE) are integrated to achieve centimeter-to-nanometer cross-scale imaging of various samples inside the narrow channel. WLE can provide a 120° wide field-of-view (FOV) to locate the area to be inspected. STABLE nano-endoscopy can provide subcellular resolution for precise analysis. Experimentally, we apply the STABLE nano-endoscopy to a number of practical sample images, including pathological samples, nanomaterials, and living mice. Our studies illustrate the potential to significantly expand the flexibility of STABLE nano-endoscopy for useful applications in the life sciences, biology science, industrial inspection, as well as clinical diagnosis.

In the way of MMF imaging to *in vivo* applications, perturbations and coupling between spatial or polarization modes, or both, are inevitable and non-negligible[31-35] (Figs. S3-S5). It has limited the MMF applications under practical movement for about decades. We introduce the STABLE concept to eliminate the influence of unpredictable couplings and fluctuations (as shown in Fig. 1A). In the STABLE process, the spatial-frequency-beacon provides fast encoded feedback in the inner loop tracking (loop 1 in Fig. 1B). The imaging sharpness acts as encoded feedback in the outer loop (loop 2 in Fig. 1B). These cooperate to track information disorder in long-haul MMFs, thereby enabling highly stable imaging. The system includes an optical module (supplementary materials, Note 2 and Fig. S2), a probe, and an electrical module as shown in Fig. 1C. The transfer functions of loops 1 and 2 are as follows:

$$T^{t+1} = T^t + (1 - G(T^t)) * T^t \quad \text{for loop 1} \quad (1)$$
$$H^{t+1} = H^t + (1 - O(H^t)) * H^t \quad \text{for loop 2}. \quad (2)$$

Here, $G$ and $O$ represent the normalised intensity of the spatial-frequency-beacon and image sharpness, respectively. $H = \exp(-id\sqrt{k_0^2 + l^2 + k^2})$ is the propagation factor. $k_0 = 2\pi/\lambda$ is the wavenumber in free space and $t$ is the index of retrieval.

Considering endoscopy applications as an example, eliminating the influence of the dynamic changes requires searching the TM based on the light reflection of the closed space immediately during imaging. In previous reports, the tracking is based on speckle measurement of the whole image and finding the correct TM after the change requires several minutes[24]. The main feature of this concept is to track and control the unpredicted coupling and deformation in spatial-frequency domain. Figure 2B shows that the mode-coupling efficiency in the low-spatial-frequency domain (0–ν/4) is significantly higher than that in the high-spatial-frequency domain. It implies that low-frequency signals are more sensitive to mode coupling than high-frequency signals. Consequently, a point around the center of the spatial-frequency domain is selected to track the dynamic changes and offer feedback signals to search for the TM in the inner loop. This is referred to as a 'spatial-frequency-beacon' (Fig. 1A). Since the spatial-frequency-beacon tracking only requires point intensity detection instead of speckle intensity measurement of the whole image, the time required to find the correct TM is approximately $10^{-3}$ s. The search speed is enhanced by a factor of $10^4$ compared to the spatial domain tracking (Fig. 2C, Table S1).

The STABLE process is verified by recovering images after deformation. The shape changes of the fibers can be classified as translation, rotation, twisting, and bending (Fig. 2A)). The number of possible deformations increase with the length of the fibers. Moreover, different deformation types can be combined to create more complex problems. Thus, MMF imaging under practical movement has been the bottleneck for about decades. Here, deformation degeneration is introduced to simple the problem significantly. Simulation and experimental results (supplementary materials, Note 5 and Movies S2, S3) suggest that the degeneration degrees of the TM induced by different deformations vary over a wide range. For instance, under rotating, limited translation and small twist of the fibers, the output between before and after deformation maintains high correlation (Figs. S13, S14). In these three cases, it is unnecessary to measure and update the TM. Simultaneously, with the increase in numerical aperture and decrease in the diameter, the degeneration degrees will be further declined (Fig. S11). Consequently, it is found that the most difficult task is to search the most suitable TM under bending and its combination. In STABLE, we pre-measured and created a high effective database containing a set of TMs and reflection matrices under different bending states. Furthermore, the spatial-frequency-beacon is generated using a reflection matrix (RM). It describes the linear input–output response of an MMF at the same port. The RM can be obtained using the Onsager reciprocity relationship and end-facet Fresnel reflection[35], as shown in Eq. (3).

$$RM = T^T * F * T \qquad (3)$$

Here, $T^T$ represents the transpose of TM. $F$ is the end-facet Fresnel reflection matrix.

The practically intriguing feature of STABLE-based endoscopy arises from the ability of stable imaging under bendable, foldable, or even twistable shapes, without detecting the distal facet. This makes the diagnosis more robust and accurate for *in vivo* imaging inside narrow lumen or in the tissues of solid organs, as desired in clinical applications. We have the MMF with a diverse range of shapes to mimic delicate but difficult-to-reach organs, such as bronchioles and coronary. As shown in Fig. 2D-E and Movie S1, images can be recovered quickly under operational movements. The detailed information on atherosclerotic and lung cell is preserved well under various shapes. These results demonstrate the potential of STABLE to provide high-definition pathological hallmarks of organs and define the potential risk for lesions.

To further verify the stability of the proposed scheme over a long duration, we used STABLE to simultaneously calibrate the optical path drift (supplementary materials, Note 4) and compensate for disorder. Over a duration of five days, we captured the different areas of the small intestine of a sheep and combined the images captured at different time (Fig. 3B). In a wide FOV image up to 1 mm image (Fig. 3, B to D), the goblet cell, absorptive cell, and central lacteal were clearly visible. They are consistent with the conventional wide-field microscope imaging (Fig. S17). For a focal point, stable focusing was achieved over one week (as shown in Fig. 3E), indicating that the system can work stably and consistently over at least one week. This is important for long-term *in situ* observations in life science. Moreover, local temperature change and liquid environment inevitably change the refractive index of the fiber and bring spherical aberration. This often distorts the true information. By using STABLE, we can obtain robust results in those complex environments (Fig. 3A). Broad bean anther, onion pollen, basswood segregation, and lily pollen were placed underwater to capture a clear fluorescent image via a 50-m MMF (Fig. 3, J to M). The growth of ice crystals in the capillary was recorded via MMF imaging (Fig. S18). Without the STABLE process, the focal plane shifts rapidly owing to the temperature change at the front of the fiber. This temperature change was well compensated by our STABLE system. The front view of the ice growth inside a 500-μm channel is clearly shown. The observation of ice growth in capillary may give some inspiration in the investigation of plant antifreeze and microchemical reaction.

To validate the stability of the method in long cavity imaging, we used a 200-m-long commercially available MMF to obtain high-contrast projection on an optical table (as shown in Fig. 3N). As the fiber length increases, slight deformations are inevitable. This change can be compensated by the STABLE. Using the same optical fiber, we performed fluorescence imaging on CdSe nanowire and fluorescent lens paper (Fig. 3O-P). This result showed that STABLE has potential applications not only in biology but also in material science.

The stable feature of such a flexible probe provides the possibility to achieve *in vivo* volumetric subcellular imaging. Volumetric imaging is difficult with the conventional MMF imaging method due to the extremely difficulties in alleviation of mode and polarization dispersion during the generation of TM stack at different axial distance (Fig. S5). In this study, the mode dispersion, polarization-mode dispersion, and mode-dependent losses occurring along the fiber length are resolved via full-vector modulation (FVM) (Eq. (4)).

$$\psi_{in}^{u,v} = T^{Tik} H(k,l) \psi_{out}^{k,l},  \qquad (4)$$

where $T^{Tik}$ is the Tikhonov regularisation of $T$.

Simulations and experiments indicate that FVM can convert the MMF function to a highly efficient programmable full-vector holographic device (Fig. S5F-I). It converts the arbitrary polarization states and spatial-frequency of the incident field into any desired polarization state and spatial-frequency. Meanwhile, FVM is combined with the pre-compensation on mode-dependent loss and polarization-mode dispersion (supplementary materials, Note 3). Thus, a high-contrast full-vector TM stack can be obtained with a single measurement. In practical applications, the combination of TM within the acquired stack with image sharpness $O$ in loop 2 can fix the preferable imaging planes within the depth range for stable volumetric imaging.

A series of 7 × 7 focal spots are generated at a distance of 200 μm from the distal facet of the fiber (Fig. S5D). Without STABLE, the power ratio—the ratio of the focal power to the total output power of the fiber—decreased significantly to less than 10% when the focal depth increased beyond 40 μm; moreover, the focal spots became approximately submerged in the noise (Fig. 4G). The decay in the power ratio was considerably mitigated by FVM. The power ratio remained at 85 % for focus depths of up to 200 μm. These results verify that STABLE significantly improves the focusing ability at high depths. Next, we performed volumetric imaging of a sample with 5-μm-diameter fluorescent spheres distributed randomly in a glue matrix. A single MMF with a core diameter of 105 μm and 0.22 NA was used to excite the fluorescent spheres and collect the fluorescence (Fig. 4H). The fluorescent images showed that the spheres could be clearly resolved at imaging depths greater than 200 μm from the fiber facet (details are given in Fig. S5E and Fig. S8). In addition, the volumetric fluorescent images could clearly distinguish the structure of the compound eye (Fig. 4I). This method can obtain volumetric imaging without mechanical moving and allows high-speed imaging by random-access scanning (in the selected regions of interest).

Based on disorder fast tracking and FVM, fluorescence emission difference can then be further utilized to break the diffraction limit to about λ/3NA (supplementary materials, Note 6). Two samples are used to verify the resolution of STABLE nano-endoscopy (USAF 1951 test target and green fluorescent microspheres of 100 nm diameter). The spot scan and difference images are shown in Figs. 4J and 4K. The intensity profiles along the line indicated by white arrows and difference images clearly show 780 nm linewidth in USAF 1951 test target (with NA = 0.22 MMF). Two nanoparticles separated by 250 nm can be distinguished (with effective NA = 0.65 MMF). This implies that the resolution of the STABLE nano-endoscopy is approximately $\lambda/3NA$, which has improved 1.5 times compared to its diffraction limit. Further, axial sub-diffraction-limited imaging can be obtained by STABLE (as shown in Fig. S20). Changing the working mode from diffraction-limited imaging to the sub-diffraction-limited mode does not require special fluorescent dye or stimulated radiation losses. This demonstrates that STABLE has the potential to obtain information about the dynamic interactions between intracellular structures in living cells.

Finally, for clinical analysis, localizing a target area in a complex scene is necessary. To bridge the gap between macroscopic and microscopic morphology, wide FOV and high-resolution 3D imaging capabilities need to be combined. We integrated the MMF with a self-developed WLE by placing the MMF in the biopsy channel of WLE. An integrated endoscope was used to observe a sample in the bronchus model and pig oesophagus (Movie S4 and Fig. 4B-F). Fluorescent markers in the outer bottom bronchial tubes were observed synchronously using the MMF and WLE. The WLE offered a wide FOV of approximately 120°, enabling global observation of the oesophagus and bronchus (Fig. S16). It also functioned as a guiding tool for the MMF probe toward the diagnosis area. The MMF could thus provide sub-micron 3D imaging of the diagnosis area. The cell is clearly seen in the MMF imaging. This tool could help doctors fast track features from different areas to diagnose benign and malignant lesions.

With STABLE, the *in vivo* high-resolution imaging of gastrointestinal tract of mice is first demonstrated. Under the guidance of WLE, MMF (NA = 0.22) enables real-time imaging of living cells following spraying topical fluorescent stains inside the tongue, oesophagus, colon, stomach and small intestine (as shown in Fig. 5). Based on the STABLE, sub-diffraction-limited imaging showing different aspects of the small intestinal mucosa is analysed at a subcellular level (as shown in Fig. 5E). In a healthy colon, epithelial nuclei surround regular, round crypt openings in transepithelial sections (as shown in Fig. 5I). Crucially, with subsurface imaging, the spatial resolution degrades due to scattering and aberrations in biological tissues (as shown in Fig. S23). This impact can be eliminated using MMF as a needle puncture to provide large-depth sampling of the submucosal (as shown in Fig. 5K and Movie S5). Compared with healthy colonic tissues, cells are specifically enriched in colon cancer tissue and the crypt structure disappeared (as shown in Fig. 5N). They are consistent with histology (Fig. 5H). Unlike other microscopy *e.g.* confocal laser endomicroscopy based on fiber bundles (as shown in Fig. S22), an important progress of the STABLE imaging setup is its ultra-capacity in ultra-thin diameter, its ability to vary the focal depth to get the optimum contrast and signal intensity and its ability to overcome the diffraction limit.

In summary, we developed a STABLE approach that enables minimally invasive, highly stable, sub-diffraction-limited, and high-contrast 3D long-fiber endoscopy in complex environments. In the developed STABLE system, the spatial-frequency-beacon involves fiber-state retrieval. It compensates the disorder and misplacement in the MMF. The FVM modulation controls the axial position of the imaging. Fluorescence emission difference is combined introduced to break the resolution to sub-diffraction limited imaging. This method paves the way for *in vivo* implementation of numerous fiber-endoscopy techniques, including polarization, super-resolution, and structured-light imaging approaches. Despite its many advantages, the proposed system has limited imaging speed owing to limited transmission bandwidth and modulation speed of modulator. This limitation could be resolved in future by improved algorithms and faster spatial light modulators. We anticipate that STABLE nano-endoscopy will provide an exciting new modality to endoscopy widely used in basic and applied research in the clinical medicine, biology and industry.

**Funding:**

National Natural Science Foundation of China (No. 62020106002, 61735017,61822510)

National Key Basic Research Program of China (No. 2021YFC2401403)

Major Scientific Research Project of Zhejiang Laboratory (No. 2019MC0AD02)

The Zhejiang University Education Foundation Global Partnership Fund

**Author contributions:**

Conceptualisation: QY, ZW, XL

Methodology: QY, ZW, ZYD

Investigation: ZW, ZYD, QY

Visualisation: ZW, ZYD, JGX

Funding acquisition: QY

Project administration: QY

Supervision: QY, XL, LQW, JBT, WC

Writing – original draft: ZW, QY

Writing – review & editing: ZW, ZYD, CLP, QY, XL, CFK, SGL

**Competing interests:** Authors declare that they have no competing interests.

**Data and materials availability:** All data are available in the main text or the supplementary materials.

**Supplementary Materials**

Materials and Methods

Supplementary Text

Figs. S1 to S23

Table S1

References (*36-49*)

Movies S1 to S5

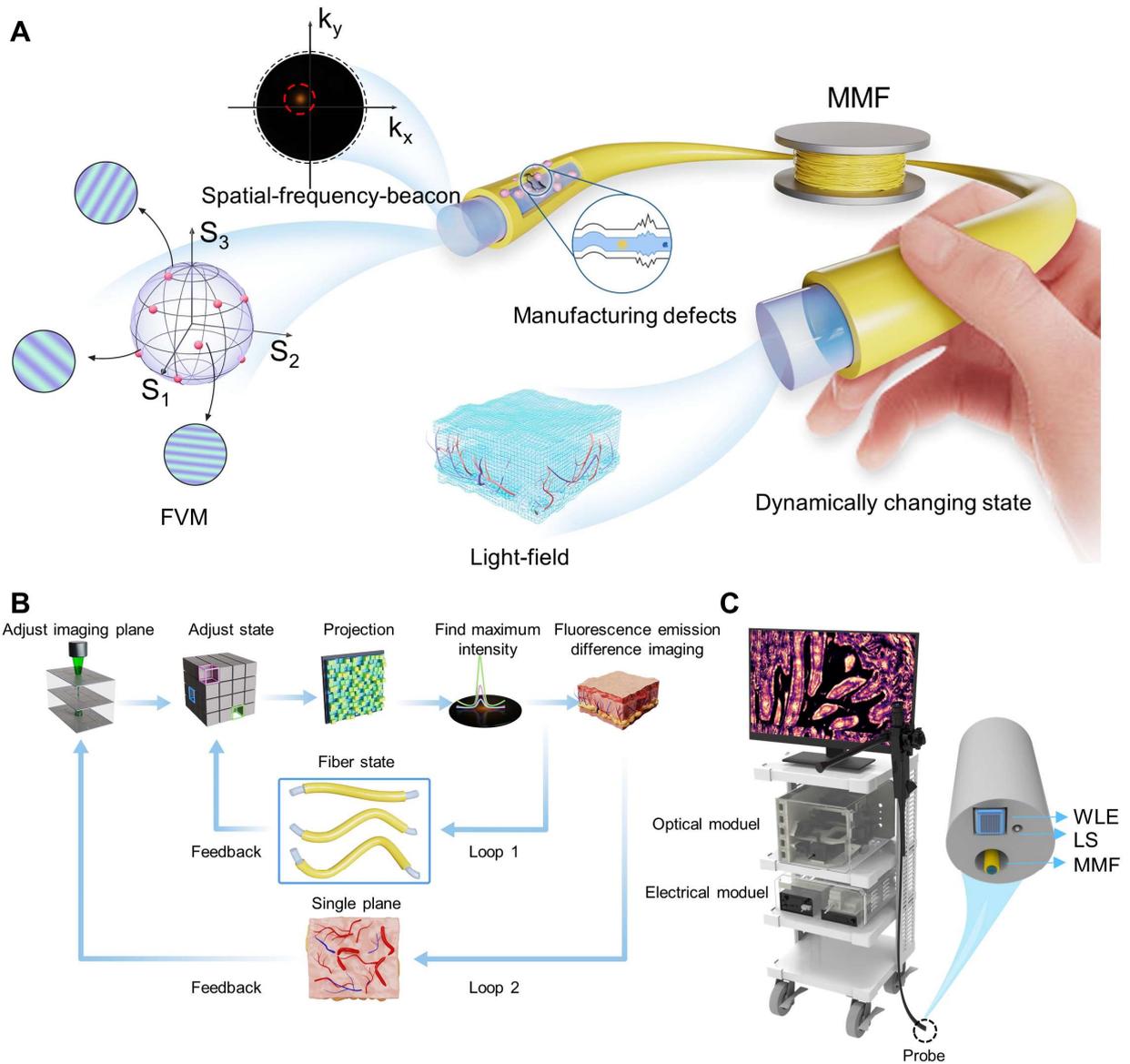

**Fig. 1. Principle of the STABLE system.** (**A**) STABLE can detect and track disorders in the MMFs introduced by practical movements and manufacturing defects, thus retaining high SNR imaging. The input light is composed of all spatial-frequency modes in different polarization states, which are spread over the Poincaré sphere. The spatial-frequency-beacon is located in the reflection spatial-frequency domain. (**B**) MMF STABLE process. Loop 1 tracks disorder, and Loop 2 tracks focal plane deviations. (**C**) Endoscope prototype. The black dotted line shows the probe. WLE: white-light endoscopy. LS: light source of WLE. MMF: multimode fiber.

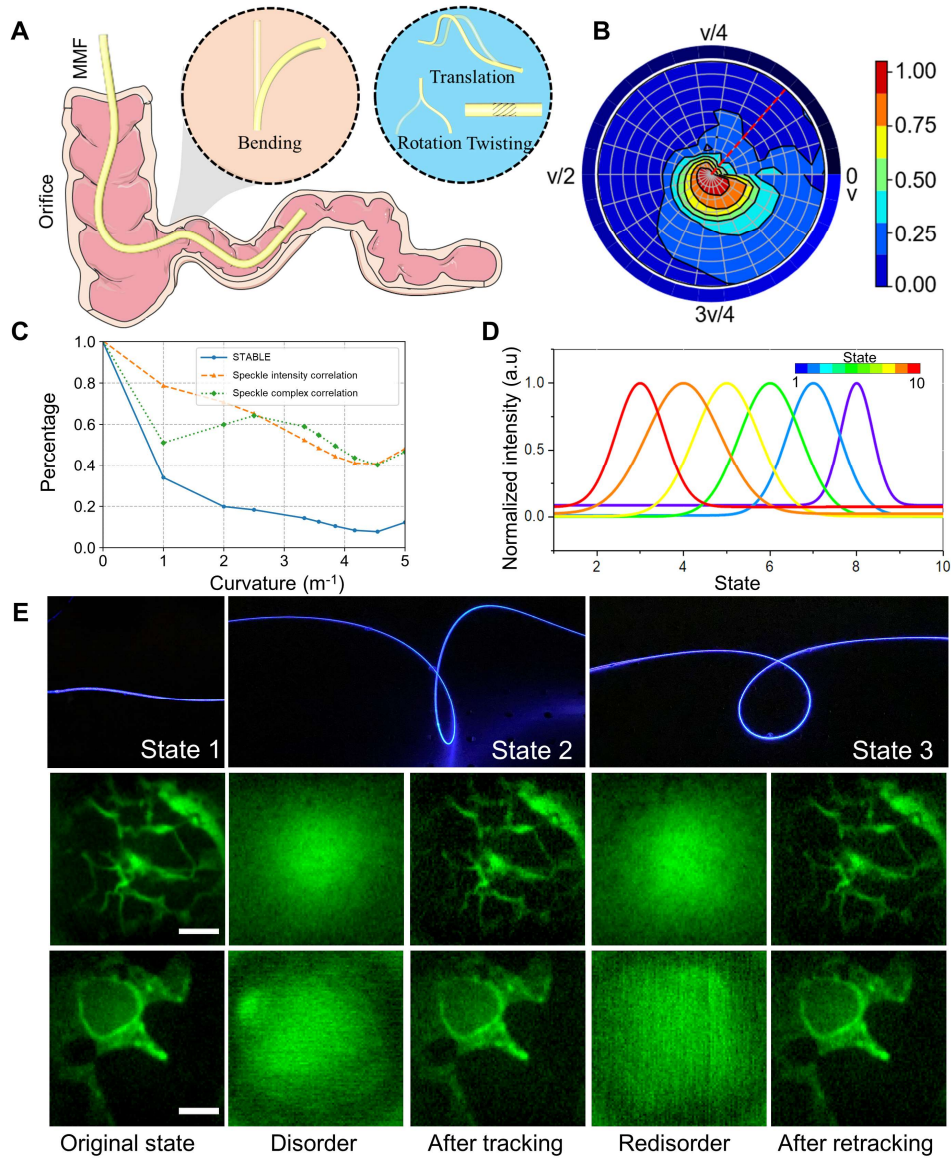

**Fig. 2. Deformation disorder track and compensation in MMF imaging.** (**A**) Types of fiber deformations. Orange circle: bending deformation. Blue circle, other three deformations: translation of a fixed curvature fiber, rotation around a central axis, and twisting deformation. (**B**) Spatial-frequency-beacon intensity variation with bending at different locations in the spatial frequency domain. The angular coordinate is spatial frequency, and v is the maximum spatial frequency of the MMF. The radial coordinate represents curvature from 0 m$^{-1}$ to 5 m$^{-1}$. The percentage represents the ratio of spatial-frequency-beacon intensity in straight state to that in different curvature. (**C**) The feedback of different probe under bending. The spatial-frequency-beacon locates at the red dashed line region of (B). (**D**) Retrieval of the bending state based on spatial-frequency-beacon tracking. (**E**) The state change and imaging recovery. The first line represents different fiber states, while the second and third lines represent the tracking and recovery process of MMF imaging of atherosclerotic and lung pathology slides under the corresponding state, respectively. Scale bar is 25 μm.

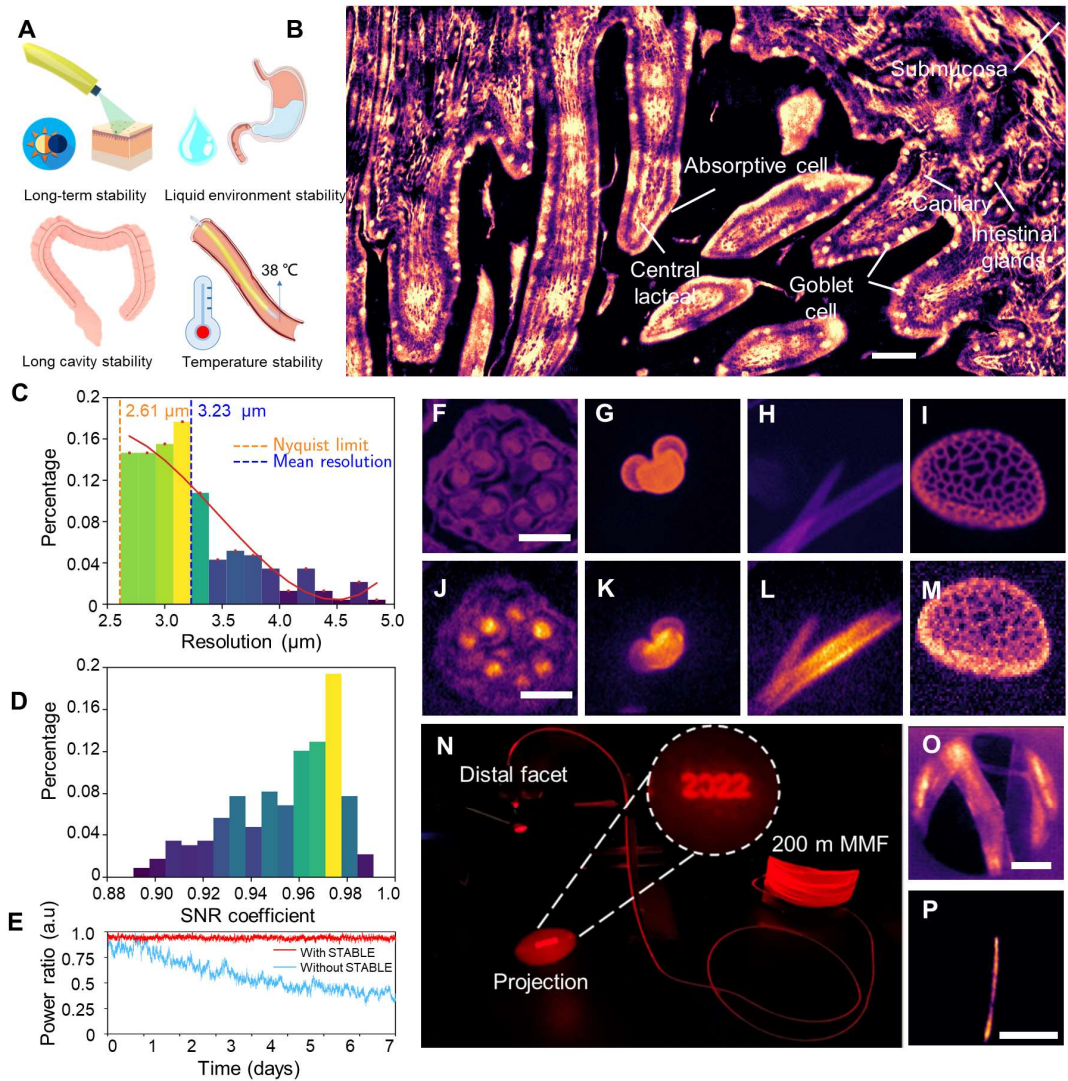

**Fig. 3. Complex environments track and compensation in MMF imaging.** (**A**) Concept diagram of endoscopy in complex environment (long-term, liquid, long cavity, temperature variations). (**B**) Image of a sheep's small intestine captured via MMF endoscopy. The types of cells are annotated on the image. The images are captured at different time during five days and can be stitching to a large FOV image without distortion, indicating the long term stability of the system. This image corresponds to the dashed box in Fig. S17A. The scale bar is 150 μm. (**C**) Resolution of each image before stitching. The horizontal axis indicates resolution of images. The vertical axis represents the percentage of total number of images. (**D**) SNR of each image before stitching. The horizontal axis indicates the percentage of total number of images. The vertical axis represents the SNR of images. (**E**) Focus intensity variation during a long period of time. (**F–I**) Wide-field microscopic images of broad bean anther, onion pollen, basswood segregation, and lily pollen samples, respectively. The scale bar is 40 μm. (**J–M**) Respective 50-m MMF images of the samples in Fig. 4, F to I placed underwater. The scale bar is 40 μm. (**N**) High contrast projection through 200-m MMF. (**O**) Imaging of fluorescent lens paper with 200-m MMF. Scale bar is 15 μm. (**P**) Imaging of CdSe nanowire with 200-m MMF. Scale bar is 15 μm.

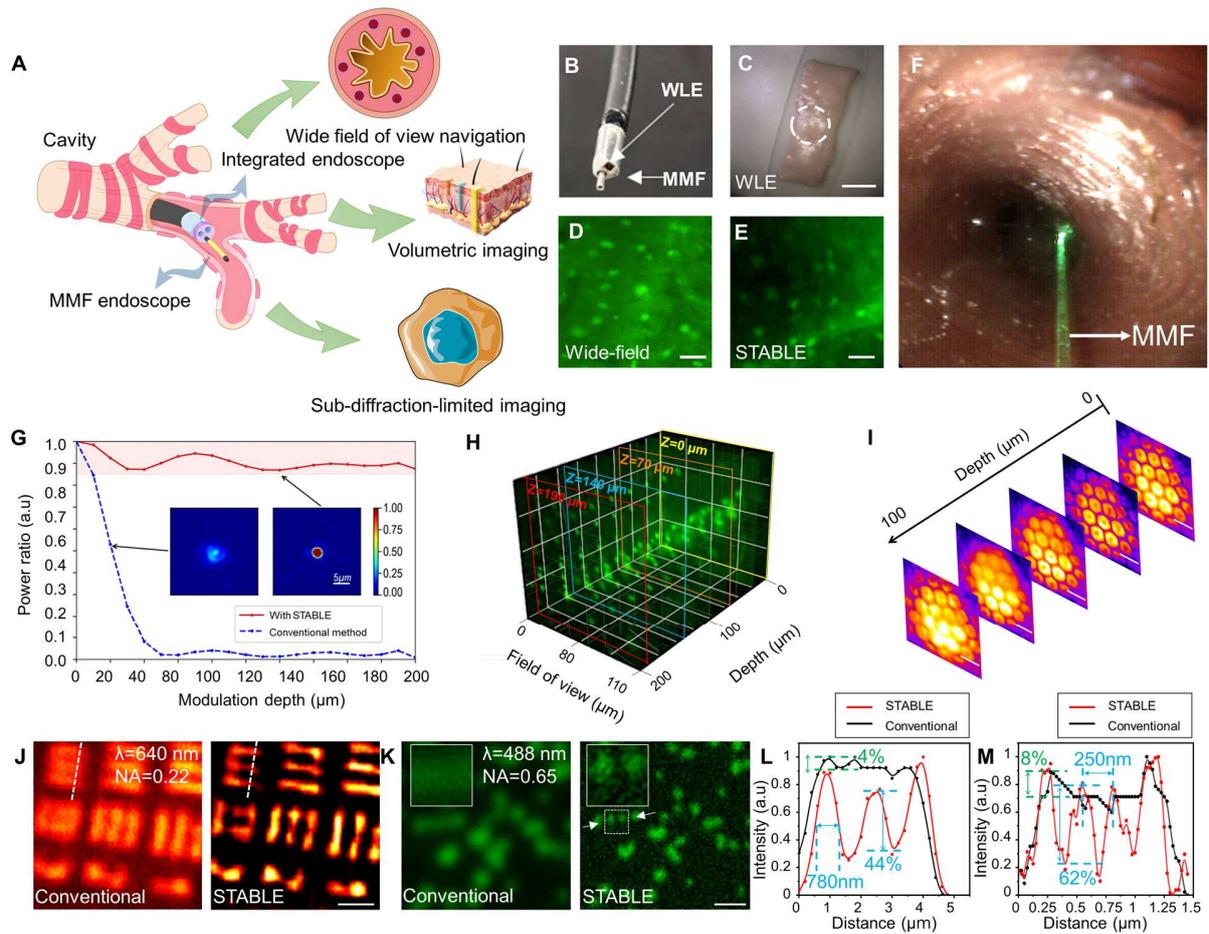

**Fig. 4. 3D, super-resolution imaging and cross-scale imaging characteristics of STABLE nano-endoscopy.** (**A**) Schematic of STABLE nano-endoscopy for ultra-stable slim endoluminal imaging. (**B**) The probe of integrated endoscope. (**C**) Pig oesophagus imaging through WLE. The scale bar is 1.8 cm. (**D**) Pig oesophagus fluorescence imaging by a wide-field microscope. The scale bar is 25 μm. (**E**) Pig oesophagus fluorescence imaging with MMFs. The scale bar is 25 μm. (**F**) Wide field of view navigation in the bronchial model. (**G**) The diagram of normalized power-ratio variation with imaging distance. (**H**) Fluorescence imaging for fluorescent microsphere at different focal planes in the glue matrix. (**I**) Butterfly compound eye was captured at different focal planes. The scale bar is 28 μm. (**J-M**) Resolution analysis of nano-endoscope. (**J**) Comparison of the MMF images of USAF 1951 test target with conventional and STABLE technique. The resolution of STABLE is 1.5 times improved. The scale bar is 3 μm. (**K**) Comparison of the MMF images of fluorescent microsphere with conventional and STABLE technique. The resolution of STABLE is 1.5 times improved. The scale bar is 0.5 μm. The mini picture is the local magnified image of the dotted box in the big picture. (**L**) Normalised intensity profiles along lines are indicated by white arrows in (J). The solid black line represents conventional imaging. The solid red line represents STABLE imaging. (**M**) Normalized intensity profiles along lines indicated by white dotted line in (K). The solid black line represents conventional imaging. The solid red line represents STABLE imaging.

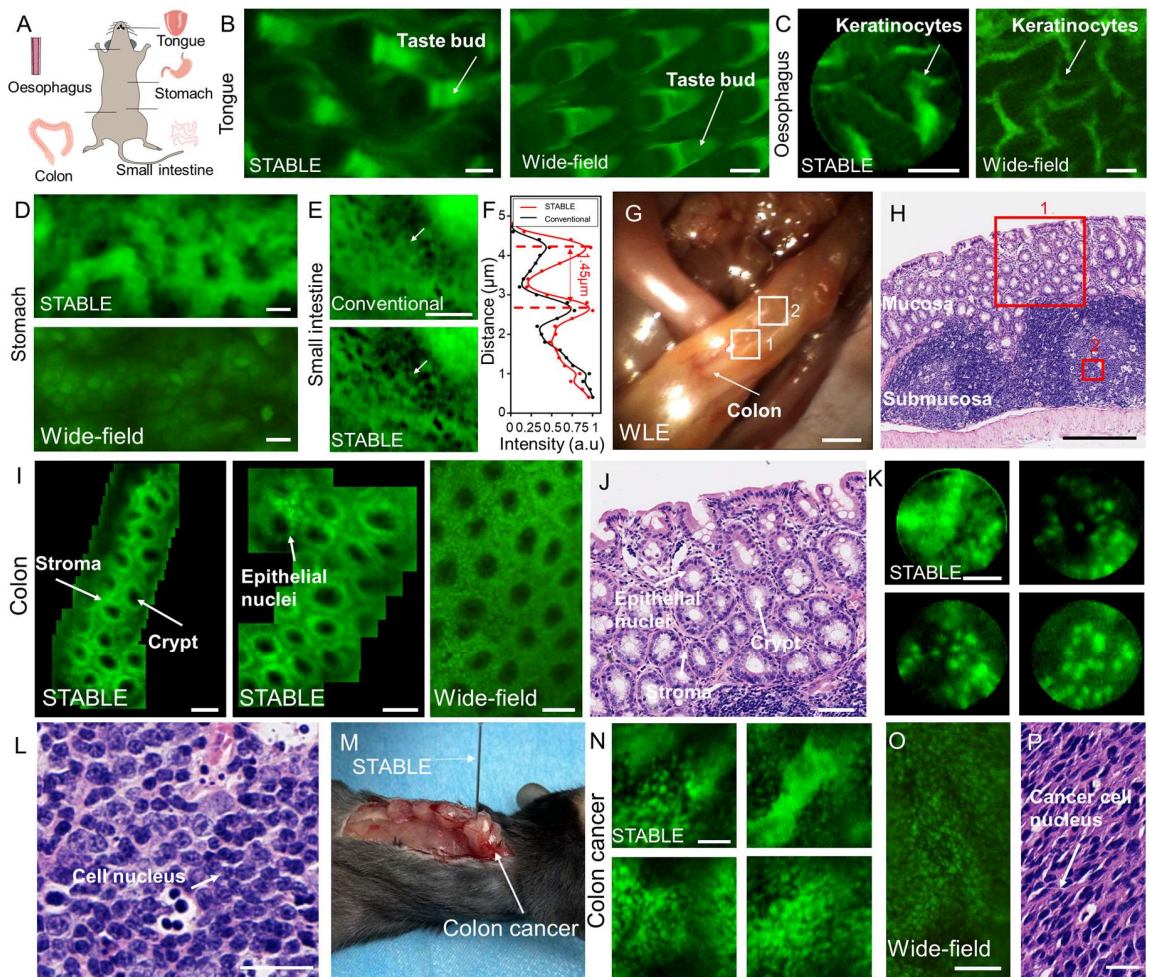

**Fig. 5. STABLE nano-endoscopy for *in vivo* gastrointestinal imaging.** The STABLE can provide resolution similar to conventional wide-field microscopy. However, the latter can't be used in the lumen due to its large size and rigid operation. (**A**) Schematic of *in vivo* imaging. (**B-C**) Stitching and individual FOV fluorescence imaging through the STABLE nano-endoscopy and wide-field microscopy at tongue and oesophagus. The scale bar is 30 μm. (**B**) Tongue. (**C**) Oesophagus. (**D**) Stitching fluorescence imaging through the STABLE nano-endoscopy and wide-field microscopy at stomach. The scale bar is 15 μm. (**E**) Sub-diffraction-limited fluorescence image of the small intestinal mucosa. The scale bar is 5 μm. (**F**) Normalised intensity profiles along lines indicated by white arrows in (E). (**G**) WLE images showing regions of healthy colon tissue. The scale bar is 4 mm. (**H**) Histology (H&E) for healthy colonic tissue is shown. The scale bar is 300 μm. (**I**) Stitching fluorescence image of the healthy colon (regions 1 and 2 in (G)) by STABLE nano-endoscopy, wide-field microscopy. The scale bar is 55 μm. (**J**) Magnified portion (mucosa) of region 1 in (H). The scale bar is 55 μm. (**K**) Images of colon tissue observed during a direct insertion of the fiber ~400 μm into the mucosa layer. The scale bar is 35 μm. (**L**) Magnified portion (submucosa) of region 2 in (H). The scale bar is 25 μm. (**M**) Experimental setup of colon cancer imaging by STABLE nano-endoscopy. (**N**) Fluorescence image of the colon cancer tissue acquired using STABLE nano-endoscopy. The scale bar is 20 μm. (**O**) Colon cancer tissue fluorescence imaging by a wide-field microscope. The scale bar is 20 μm. (**P**) Histology (H&E) for colon cancer tissue is shown. The scale bar is 20 μm.